\documentclass[aps,twocolumn,showpacs,preprintnumbers,amsmath,amssymb,superscriptaddress,floatfix,nofootinbib]{revtex4-1}

\usepackage[utf8]{inputenc}
\usepackage{slashed}
\usepackage{soul}
\usepackage{mathrsfs,bm}
\usepackage{txfonts}
\usepackage{amssymb}
\usepackage{indentfirst}
\usepackage{graphicx,booktabs}
\usepackage{multirow}
\usepackage{overpic}
\usepackage{color}
\usepackage{amssymb}
\usepackage{hyperref}

\begin{document}
\title{Emergence of a complete heavy-quark spin symmetry multiplet:\\
  seven molecular pentaquarks in light of the latest LHCb analysis }
\author{Ming-Zhu Liu}

\author{Ya-Wen Pan}
\author{Fang-Zheng Peng}
\affiliation{School of Physics and
Nuclear Energy Engineering,  Beihang University, Beijing 100191, China}

\author{Mario S\'anchez S\'anchez}
\affiliation{Centre d'\'Etudes Nucl\'eaires, CNRS/IN2P3, Universit\'e de Bordeaux, 33175 Gradignan, France}

\author{Li-Sheng Geng}\email{lisheng.geng@buaa.edu.cn}

\affiliation{School of Physics and
Nuclear Energy Engineering,  Beihang University, Beijing 100191, China}
\affiliation{
  Beijing Key Laboratory of Advanced Nuclear Materials and Physics \&Beijing Advanced Innovation Center for Big Data-based Precision Medicine, Beihang University, Beijing, 100191,China
  }

\author{Atsushi Hosaka}\email{hosaka@rcnp.osaka-u.ac.jp}
\affiliation{Research Center for Nuclear Physics (RCNP), Osaka University, Ibaraki 567-0047, Japan}
\affiliation{Advanced Science Research Center, Japan Atomic Energy Agency (JAEA), Tokai 319-1195, Japan}

\author{Manuel Pavon Valderrama}\email{mpavon@buaa.edu.cn}

\affiliation{School of Physics and
Nuclear Energy Engineering,  Beihang University, Beijing 100191, China}
\affiliation{International Research Center for Nuclei and Particles
  in the Cosmos \&
  Beijing Key Laboratory of Advanced Nuclear Materials and Physics,
  Beihang University, Beijing 100191, China}

\date{\today}

\begin{abstract}
  A recent analysis by the LHCb collaboration suggests the existence of 
  three narrow pentaquark-like states --- the $P_c(4312)$,
  $P_c(4440)$ and $P_c(4457)$ --- instead of just one
  in the previous analysis (the $P_c(4450)$).
  The closeness of the $P_c(4312)$ to the $\bar{D} \Sigma_c$ threshold
  and the $P_c(4440)$/$P_c(4457)$ to the $\bar{D}^* \Sigma_c$ one
  suggests a molecular interpretation of these resonances.
  We show that these three pentaquark-like resonances can be naturally
  accommodated in a contact-range effective field theory description
  that incorporates heavy-quark spin symmetry.
  This description leads to the prediction of all the seven possible
  S-wave heavy antimeson-baryon molecules (that is, there should be four
  additional molecular pentaquarks in addition to the $P_c(4312)$, $P_c(4440)$
  and $P_c(4457)$), providing the first example of a heavy-quark spin
  symmetry molecular multiplet that is complete.
  If this is confirmed, it will not only give us an impressive example of
  the application of heavy-quark symmetries and effective field theories
  in hadron physics: it will also uncover a clear and powerful
  ordering principle for the molecular spectrum, reminiscent of
  the SU(3)-flavor multiplets to which
  the light hadron spectrum conforms.
\end{abstract}

\maketitle

In 2015 the LHCb collaboration discovered the existence of
two pentaquark-like resonances, which are usually referred to as
$P_c(4380)$ and $P_c(4450)$ due to their masses~\cite{Aaij:2015tga}.
This experimental discovery triggered intense theoretical speculations
on the nature of these states, their decays and production mechanisms.
In particular the closeness of the $P_c(4450)$ to a few meson-baryon thresholds
($\bar{D}^* \Sigma_c$, $\bar{D}^* \Sigma_c^*$,
$\bar{D} \Lambda_{c}(2595)$ and ${\chi_{c1}p}$)
leads naturally to the conjecture that it is a meson-baryon bound state
{(a conjecture further cemented by a series of theoretical predictions
that predated its
observation~\cite{Wu:2010jy,Wu:2010vk,Wu:2010rv,Xiao:2013yca,Karliner:2015ina,Wang:2011rga,Yang:2011wz})},
with the most popular explanations being
a $\bar{D}^* \Sigma_c$~\cite{Roca:2015dva,He:2015cea,Xiao:2015fia} or
a $\bar{D}^* \Sigma_c^*$ molecule~\cite{Chen:2015loa,Chen:2015moa}
(in these two cases in the isospin $I=\tfrac{1}{2}$ configuration and
probably with a small admixture of
$\bar{D} \Lambda_{c}(2595)$~\cite{Burns:2015dwa,Geng:2017hxc}),
and a {${\chi_{c1}p}$} molecule~\cite{Meissner:2015mza}.
There are also non-molecular explanations for this state, which include
that it might be a genuine pentaquark~\cite{Diakonov:1997mm,Jaffe:2003sg,Yuan:2012wz,Maiani:2015vwa,Lebed:2015tna,Li:2015gta,Wang:2015epa},
that threshold effects might play a role~\cite{Guo:2015umn,Liu:2015fea}
(see also Ref.~\cite{Bayar:2016ftu} for a detailed discussion),
baryocharmonia~\cite{Kubarovsky:2015aaa}, a molecule
bound by {\it colour chemistry}~\cite{Mironov:2015ica}
and a soliton~\cite{Scoccola:2015nia}.

{The original analysis of Ref.~\cite{Aaij:2015tga} has been recently updated
by the LHCb collaboration in Ref.~\cite{Aaij:2019vzc}, where it has been
found that} the previous $P_c(4450)$ actually contains two peaks
--- the $P_c(4440)$ and $P_c(4457)$ ---
and that there is a third narrow peak, the $P_c(4312)$.
Their masses and widths are
\begin{eqnarray*}
  m_{P_{c1}}&=&4311.9\pm 0.7^{+6.8}_{-0.6}, \quad
  \Gamma_{P_{c1}}=9.8\pm2.7^{+3.7}_{-4.5}, \nonumber \\
  m_{P_{c2}}&=&4440.3\pm 1.3^{+4.1}_{-4.7}, \quad
  \Gamma_{P_{c2}}=20.6\pm4.9^{+8.7}_{-10.1}, \nonumber \\
  m_{P_{c3}}&=&4457.3\pm 0.6{}^{+4.1}_{-1.7}, \quad
  \Gamma_{P_{c3}}=6.4\pm2.0^{+5.7}_{-1.9},  \nonumber 
\end{eqnarray*}
all in units of ${\rm MeV}$ and for which we have used the notation
$P_{c1}$, $P_{c2}$ and $P_{c3}$ to refer to the three states $P_c(4312)$,
$P_c(4440)$ and $P_c(4457)$ (that is, we have ordered
them according to their masses).
It is interesting to notice that the mass of the previous $P_c(4450)$ roughly
coincides with the geometric mean of the masses of the new
$P_c(4440)$ and $P_c(4457)$.
The $P_c(4312)$ pentaquark-like state is near to the $\bar{D} \Sigma_c$
threshold, while the other two are close to the $\bar{D}^* \Sigma_c$ one.
When translated into binding energies we obtain $B_1 = 8.9$, $B_2 = 21.8$
and $B_3 = 4.8\,{\rm MeV}$ for the $P_c(4312)$, $P_c(4440)$
and $P_c(4457)$, respectively.
Of course this closeness to threshold has already been noted
by theoreticians in Refs.~\cite{Chen:2019bip,Chen:2019asm,Guo:2019kdc}.
If these findings are confirmed it will not only strongly support
the molecular hypothesis, but it will also provide us with
the most impressive illustration of the application of
heavy-quark spin symmetry (HQSS)~\cite{Isgur:1989vq,Isgur:1989ed,Neubert:1993mb,Manohar:2000dt} to hadronic molecules so far.
In particular this experimental analysis will result
in the prediction of the first full HQSS molecular multiplet
of the hidden-charm molecular pentaquarks.

Heavy-hadron molecules, i.e. bound states that include one or
more heavy hadrons, were conjectured decades
ago~\cite{Voloshin:1976ap,DeRujula:1976qd}.
Owing to the combination of light- and heavy-quark content,
heavy-hadron molecules have a high degree of symmetry
which can be exploited to determine their spectrum~\cite{AlFiky:2005jd,Guo:2009id,Voloshin:2011qa,Mehen:2011yh,Valderrama:2012jv,Nieves:2012tt,HidalgoDuque:2012pq,Guo:2013sya,Guo:2013xga,Lu:2017dvm}.
HQSS manifests in the existence of interesting patterns
in the heavy-molecular spectrum.
The most evident of these patterns applies to the $Z_c$'s and $Z_b$'s twin
resonances discovered by BESIII~\cite{Ablikim:2013mio, Liu:2013dau,Ablikim:2013wzq, Ablikim:2014dxl} and Belle~\cite{Belle:2011aa,Garmash:2014dhx},
respectively.
If they are bound states of a heavy meson and antimeson, being either a
charm meson or bottom meson, HQSS predicts that the S-wave potential
in the $J^{PC} = 1^{+-}$ channel is~\cite{Voloshin:2011qa,Mehen:2011yh}
\begin{eqnarray}
  V(1^{+-}, P^* \bar{P}) = V(1^{+-}, P^* \bar{P}^*) \, ,
\end{eqnarray}
independently of the particle content,
where $P = D, \bar{B}$ and $P^* = D^*, \bar{B}^*$.
This specific pattern indeed explains why the $Z_c$'s and $Z_b$'s appear
in pairs, both of which are at similar distances from the $P^* P$ and
$P^* P^*$ open heavy flavour thresholds.
A similar pattern applies to the $1^{++}$ and $2^{++}$ heavy meson-antimeson
interaction~\cite{Valderrama:2012jv,Nieves:2012tt,HidalgoDuque:2012pq}
\begin{eqnarray}
  V(1^{++}, P^* \bar{P}) = V(2^{++}, P^* \bar{P}^*) \, .
\end{eqnarray}
If we assume the $X(3872)$ resonance to be a $D^* \bar{D}$ bound state with
$J^{PC} = 1^{++}$, this symmetry relation suggests the existence of a $2^{++}$
$D^* \bar{D}^*$ partner with a mass of $4012\,{\rm MeV}$.
However the location of the $X(3872)$ overlaps with the $D^{0*} \bar{D}^0$
threshold within experimental errors, which implies that the existence of
the $2^{++}$ partner of the $X$ is not guaranteed
if we take into account this error source (besides, there are other
uncertainties~\cite{Cincioglu:2016fkm,Baru:2016iwj}),
see Ref.~\cite{Liu:2019stu} for a more complete discussion.
At this point we notice that there are six possible heavy meson-antimeson
molecules, forming a HQSS multiplet that can accommodate
up to six resonances.
However the known heavy meson-antimeson molecules are all
too close to threshold, indicating that most probably
this multiplet structure is unlikely to be fully
realized in nature,
leaving us with an incomplete pattern.

{This manuscript argues that the new LHCb results~\cite{Aaij:2019vzc}
imply that the heavy antimeson-baryon molecules will probably provide
the first example of a full and intact HQSS molecular multiplet.}
For this we begin by explaining the constraints that HQSS imposes
on the S-wave heavy antimeson-baryon interaction, as has been
recently derived in Ref.~\cite{Liu:2018zzu}.
HQSS implies that we can describe the seven S-wave $\bar{D}^{(*)} \Sigma_c^{(*)}$
molecules with two coupling constants.
If we additionally assume that the heavy antimeson-baryon molecules
can be described in terms of a contact-range effective field theory
(EFT), the potential for the $\bar{D} \Sigma_c$, $\bar{D} \Sigma_c^*$,
$\bar{D}^* \Sigma_c$ and $\bar{D}^* \Sigma_c^*$ molecules
is~\cite{Liu:2018zzu}
\begin{eqnarray}
  V(\tfrac{1}{2}^{-}, \bar{D} \Sigma_c) &=& C_a \, , \label{eq:V1} \\
  V(\tfrac{3}{2}^{-}, \bar{D} \Sigma_c^*) &=& C_a \, , \\
  V(\tfrac{1}{2}^{-}, \bar{D}^* \Sigma_c) &=& C_a - \tfrac{4}{3}\,C_b\, , \\
  V(\tfrac{3}{2}^{-}, \bar{D}^* \Sigma_c) &=& C_a + \tfrac{2}{3}\,C_b\, , \\
  V(\tfrac{1}{2}^{-}, \bar{D}^* \Sigma_c^*) &=& C_a - \tfrac{5}{3}\,C_b\, , \\
  V(\tfrac{3}{2}^{-}, \bar{D}^* \Sigma_c^*) &=& C_a - \tfrac{2}{3}\,C_b\, , \\
  V(\tfrac{5}{2}^{-}, \bar{D}^* \Sigma_c^*) &=& C_a + C_b\, , \label{eq:V7} 
\end{eqnarray}
with $C_a$ and $C_b$ unknown coupling constants.
This potential is renormalized by including a separable regulator
and a cutoff $\Lambda$ in momentum space and allowing
the couplings to depend on this cutoff
\begin{eqnarray}
  \langle p | V_{\Lambda} | p' \rangle =  C(\Lambda)\,
  f(\tfrac{p}{\Lambda})\,f(\tfrac{p'}{\Lambda}) \, ,
    \label{eq:sep-regul}
\end{eqnarray}
where $p$, $p'$ are the initial and final center-of-mass momenta of the
two-body system and $C$ represents the linear combination of
$C_a$ and $C_b$ corresponding
to the heavy antimeson-baryon molecule under consideration,
see Eqs.~(\ref{eq:V1}-\ref{eq:V7}) for details.
For the regulator we choose a Gaussian one, $f(x) = e^{-x^2}$,
while for the cutoff we consider the range $\Lambda = 0.5-1.0\,{\rm GeV}$,
where we notice that if the problem has been properly renormalized
the dependence of the predictions on the cutoff will be small.
The potential is then included in a dynamical equation,
e.g. Lippmann-Schwinger:
\begin{eqnarray}
  \phi(k) + \int \frac{d^3 p}{(2\pi)^3}\,\langle k | V_{\Lambda} | p \rangle
  \,\frac{\phi(p)}{B_2 + \frac{p^2}{2 \mu}} = 0 \, ,
\end{eqnarray}
with $\phi$ the vertex function, $\mu$ the reduced mass of the system
and $B_2$ the binding energy, where solutions of this dynamical
equation correspond to bound states.
Alternatively, {with the purpose of checking regulator independence},
we can use a delta-shell regulator in coordinate space
\begin{eqnarray}
  V(r; R_c) =  C(R_c)\,\frac{\delta(r- R_c)}{4 \pi R_c^2} \, ,
  \label{eq:delta-shell}
\end{eqnarray}
with a cutoff in the range $R_c = 0.5-1.0\,{\rm fm}$, i.e. of the order
of the typical hadron size, and solve the S-wave Schr\"odinger equation.
The delta-shell regulator is convenient --- it allows for analytic
results --- but we stress that any other choice of regulator will work too.
We will not show detailed results for the delta-shell or other regulators here,
but simply comment that the differences with the Gaussian regulator
are minor.

We notice that both the $P_c(4440)$ and $P_c(4457)$ are good
$\bar{D}^* \Sigma_c$ molecular candidates, but their $J^P$ is not known.
The natural expectation in hadron physics is that states with higher spin
will have a higher mass, but hadronic molecules might deviate
from this trend depending on the binding mechanism (e.g. the spectrum
predicted in Ref.~\cite{Karliner:2015ina} for the molecular pentaquarks).
Here we will not make {\it a priori} assumptions and will let
the comparison between theory and experiment decide instead.
Thus we distinguish two scenarios, $A$ and $B$, where $A$ corresponds to
assuming that the $P_c(4440)$ and $P_c(4457)$ are $J^P = \tfrac{1}{2}^{-}$
and $J^P = \tfrac{3}{2}^{-}$ respectively, while $B$ corresponds to the
opposite identification.
Each of these choices completely fixes the EFT potential and allows us
to predict the location of the $J^P =\frac{1}{2}^{-}$ $\bar{D} \Sigma_c$
molecule, which in scenario $A$ we predict at
\begin{eqnarray}
  M_A (\bar{D} \Sigma_c) = (4311.8 - 4313.0)\,{\rm MeV} \, ,
\end{eqnarray}
which is extremely close to the experimental value,
$M_{P_{c1}} = 4311.9 \,{\rm MeV}$, and where the range corresponds
to the cutoff variation $\Lambda = 0.5-1.0\,{\rm GeV}$.
In contrast to this, in scenario $B$ the $P_c(4312)$ resonance is predicted at
\begin{eqnarray}
  M_B (\bar{D} \Sigma_c) = (4306.3 - 4307.7)\,{\rm MeV} \, , 
\end{eqnarray}
which is not too close to the expected location of
the $P_c(4312)$ resonance but still compatible
within the experimental errors.
The predictions are fairly independent not only on the cutoff, but also
on the choice of regulator: had we used the delta-shell regulator of
Eq.~(\ref{eq:delta-shell}) instead of the Gaussian regulator of
Eq.~(\ref{eq:sep-regul}), the predictions would have been
\begin{eqnarray}
M_A'(\bar{D} \Sigma_c) &=& (4312.1-4313.1)\,{\rm MeV} \, , \\
M_B'(\bar{D} \Sigma_c) &=& (4306.7-4308.0)\,{\rm MeV} \, ,
\end{eqnarray}
for the $R_c = 0.5-1.0\,{\rm fm}$ cutoff range, 
  which indicates a preference for scenario $A$. This is also the case
  for other regulators, e.g. a square-well or a Gaussian
  potential in coordinate space.
  However, note that we have not propagated the uncertainty in the masses of
  the input data --- the $P_c(4440)$ and $P_c(4457)$ --- neither have we taken
  into account the experimental uncertainty in the location of
  the $P_c(4312)$. That is, the preference for scenario $A$
  is probably not particularly strong.
  For comparison purposes,
  the seminal works of Refs.~\cite{Wu:2010jy,Wu:2010vk,Wu:2010rv,Xiao:2013yca}
  predict the $J^P = \tfrac{1}{2}^-$ and
  $\tfrac{3}{2}^-$ $\bar{D}^* \Sigma_c$ states to be degenerate.
  The molecular identifications of the $P_c(4440)$ and $P_c(4457)$ that
  are derived from the one boson exchange models of
  Refs.~\cite{Chen:2019asm,He:2019ify}
  are equivalent to our scenario $A$,
  i.e. the scenario favored by our calculations.
  On the other hand,
  Ref.~\cite{Karliner:2015ina} predicts the $\tfrac{3}{2}^{-}$
  $\bar{D}^* \Sigma_c$ molecule to be more bound, a conclusion which
  is deduced from the strength and sign of the one pion exchange potential
  in S-waves and which corresponds to our scenario $B$.

\begin{table}[!ttt]
\begin{tabular}{|ccccc|}
\hline \hline
  Scenario & Molecule  & $J^{P}$ & B (MeV) & M (MeV) \\
  \hline
  $A$ & $\bar{D} \Sigma_c$ & $\frac{1}{2}^-$ & $7.8-9.0$ & $4311.8-4313.0$ \\ \hline
  $A$ & $\bar{D} \Sigma_c^*$ & $\frac{3}{2}^-$ & $8.3-9.2$ & $4376.1-4377.0$ \\ \hline
  $A$ & $\bar{D}^* \Sigma_c$ & $\frac{1}{2}^-$ & Input &$4440.3$ \\
  $A$ & $\bar{D}^* \Sigma_c$ & $\frac{3}{2}^-$ 
  & Input & $4457.3$ \\
  \hline
  $A$ & $\bar{D}^* \Sigma_c^*$ & $\frac{1}{2}^-$ 
  & $25.7-26.5$ & $4500.2-4501.0$
  \\
  $A$ & $\bar{D}^* \Sigma_c^*$ & $\frac{3}{2}^-$ & 
  $15.9-16.1$ & $4510.6-4510.8$
  \\
  $A$ & $\bar{D}^* \Sigma_c^*$ & $\frac{5}{2}^-$ & $3.2-3.5$ &
  $4523.3-4523.6$ \\
  \hline \hline 
  $B$ & $\bar{D} \Sigma_c$ & $\frac{1}{2}^-$ & $13.1-14.5$ &
  $4306.3-4307.7$\\ \hline
  $B$ & $\bar{D} \Sigma_c^*$ & $\frac{3}{2}^-$ & $13.6-14.8$ &
  $4370.5-4371.7$ \\ \hline
  $B$ & $\bar{D}^* \Sigma_c$ & $\frac{1}{2}^-$  & Input & $4457.3$ \\
  $B$ & $\bar{D}^* \Sigma_c$ & $\frac{3}{2}^-$  & Input &$4440.3$ \\
  \hline
  $B$ & $\bar{D}^* \Sigma_c^*$ & $\frac{1}{2}^-$ & $3.1-3.5$ & $4523.2-4523.6$
  \\
  $B$ & $\bar{D}^* \Sigma_c^*$ & $\frac{3}{2}^-$ & $10.1-10.2$ & $4516.5-4516.6$
  \\
  $B$ & $\bar{D}^* \Sigma_c^*$ & $\frac{5}{2}^-$ & $25.7-26.5$ &
  $4500.2-4501.0$ \\
  \hline \hline
\end{tabular}
\caption{
  Predictions for the S-wave HQSS molecular multiplet of
  heavy antimeson-baryon molecules, as derived from the 
  lowest{-}order contact-range potential which
  contains two unknown couplings $C_a$ and $C_b$.
  The potential for each particle and spin channel
  (the ``Molecule'' and ``$J^P$'' columns) can be checked
  in Eqs.~(\ref{eq:V1}-\ref{eq:V7}).
  In all cases we assume that the isospin of the listed
  molecules is $I=\tfrac{1}{2}$.
  We determine the value of the $C_a$ and $C_b$ couplings from the condition
  of reproducing the location of the $P_c(4440)$ and $P_c(4457)$ resonances,
  which are known to be close to the $\bar{D}^* \Sigma_c$ threshold.
  We do not know however the quantum numbers of the $P_c(4440)$ and $P_c(4457)$,
  but consider two possibilities instead: in scenario $A$ the $\frac{1}{2}^-$
  molecule is identified with the $P_c(4440)$ and the $\frac{3}{2}^-$ with the
  $P_c(4457)$, while scenario $B$ assumes the opposite identification. 
}
\label{tab:penta}
\end{table}

Yet, as previously explained, the really exciting aspect of being able
to determine both $C_a$ and $C_b$ is that now we can predict all
the seven heavy antimeson-baryon molecules.
This is done in Table \ref{tab:penta} for scenarios $A$ and $B$, where in both
cases the seven molecules are always predicted but the
specifics of their location changes slightly
depending on the chosen scenario, particularly
in what regards the $\bar{D}^* \Sigma_c^*$ molecules:
for scenario $A$ binding decreases with the spin quantum number,
while the contrary is true for scenario $B$.
{
We notice the prediction of a $\bar{D} \Sigma_c^*$ bound state
at $4370-4380\,{\rm MeV}$, thought the identification
with the $P_c(4380)$ pentaquark peak of Ref.~\cite{Aaij:2015tga}
is problematic owing to the broad nature of this state.}
{ 
Our conclusion that the HQSS multiplet for the heavy meson-baryon molecules
is complete has been independently confirmed in Ref.~\cite{Xiao:2019aya}.
}

To summarize, the recent analysis of the LHCb collaboration supports
the hypothesis that the pentaquark-like $P_c(4312)$, $P_c(4440)$
and $P_c(4457)$ resonances are indeed $\bar{D} \Sigma_c$ and
$\bar{D}^* \Sigma_c$ molecules.
In addition, this experimental observation unlocks the possibility of
the theoretical prediction of all the seven S-wave heavy antimeson-baryon
molecules, which incidentally provides the first example of a full
and complete HQSS multiplet for hadronic molecules.
{The identification of the $P_c(4440)$ and $P_c(4457)$
  with $\bar{D}^* \Sigma_c$ bound states is ambiguous:
  both the $J^P = \tfrac{1}{2}^-$ and $J^P = \tfrac{3}{2}^-$
  quantum numbers are in principle possible.}
Even though the specific spin of the
$P_c(4440)$ and $P_c(4457)$ molecular candidates is inconsequential
for the prediction of the HQSS multiplet, the spectroscopic
predictions of the contact-range effective field theory we use in this
manuscript indicates a preference for identifying the $P_c(4440)$ and
$P_c(4457)$ with the $J^P=\frac12^-$ and the $J^P=\frac32^-$
$\bar{D}^* \Sigma_c$ molecules, respectively.
Though the present theoretical exploration focuses only on the spectroscopy
of the molecular pentaquarks, the eventual discovery of the missing members
of the HQSS multiplets at their predicted locations will by itself
represent a very strong case in favor of their molecular nature.
Yet future investigation of their decays and production mechanisms
will be essential to disentangle the nature of
these pentaquark-like states.
Finally we stress that the idea of HQSS multiplets
provides a clear and concise ordering principle
for molecular states which, in analogy to
the SU(3)-flavour multiplets in the light
hadron sector, has the potential to help to interpret
the results of future experimental searches of exotic states
and to improve our understanding of the non-perturbative strong interaction.

\section*{Acknowledgments}

This work is partly supported by the National Natural Science Foundation of
China under Grant No. 11735003, the Fundamental Research Funds for the Central
Universities and the Thousand Talents Plan for Young Professionals.
AH acknowledges support from  Grants-in-Aid for Scientific Research
(No. JP17K05441 (C)) and for Scientific Research
on Innovative Areas (No. 18H05407).


%

\end{document}